# Revisiting the Electrified Pt(111)/Water Interfaces through an Affordable Double-Reference ab-initio Approach

Jack Jon Hinsch[1], Assil Bouzid[2], Jordan Clive Barker[1], Jessica Jein White[1], Fabien Mortier[2], Huijun Zhao[1], Yun Wang[1*]

1: Centre for Catalysis and Clean Energy, School of Environment and Science, Griffith University, Gold Coast, QLD 4222, Australia

2: Institut de Recherche sur les Céramiques (IRCER), UMR CNRS 7315-Université de Limoges, France

* Corresponding Author

E-mail: yun.wang@griffith.edu.au

**Abstract**

The electrified solid-liquid interface plays an essential role in many renewable energy-related applications, including hydrogen production and utilization. Limitations in computational modelling of the electrified solid-liquid interface have held back the understanding of its properties at the atomic-scale level. In this study, we applied the grand canonical density functional theory (GC-DFT) combined with a hybrid implicit/explicit solvation model to reinvestigate the widely studied electrified platinum-water interface affordably. The calculated double layer capacitances of the Pt(111)-water interface over the applied bias potential closely match the experimental and previous theoretical data from expensive *ab-initio* molecular dynamics simulations. The structural analysis of the interface models reveals that the applied bias potential can significantly affect the Pt(111)-water atomic interface configurations. The orientation of the water molecules next to the Pt(111) surface is vital for correctly describing the potential of zero charge (PZC) and capacitance. Additionally, the GC-DFT results confirm that the absorption of the hydrogen atom under applied bias potential can significantly affect the electrified interfacial properties. The presented affordable GC-DFT approach, therefore, offers an efficient and accurate means to enhance the understanding of electrified solid-liquid interfaces.

## 1. Introduction

Electrified solid-liquid interfaces are of great importance in many energy and electrochemistry applications, including electrocatalysis, batteries, supercapacitors, solar cells, and fuel cells [1-3]. The interface often determines the charge transfer rate of the electrochemical redox reactions, which is overseen by the applied bias electrode potential [4]. However, understanding the atomistic properties of the electrified solid-liquid interface and the thermodynamic state of interfacial processes is still challenging [5, 6]. The difficulty arises from the operando characterizations on the solid-liquid interfaces [7]. To address these issues, atomistic modelling is a powerful and complementary tool to access the structural and electronic properties of the electrified interface [8].

Several *ab-initio* simulation strategies have been developed thus far to characterize the electrified solid-liquid interface [8]. The computational hydrogen electrode (CHE) model is one of the widely used approaches for studying the electrified solid-liquid interface [9, 10]. In this method, the standard hydrogen electrode (SHE) serves as the reference potential. A pair of protons ($H^+$) and electrons ($e^-$) are introduced in each elementary step to charge transfer processes [11-13]. However, this method dismisses the influence of the applied bias potential on the properties of the solid-liquid interface [14]. To accurately reveal the dynamic nature of the electrified solid-water interface, the *ab-initio* molecular dynamics (AIMD) method has been employed by several groups to investigate the Pt(111)-water interface, which is the most studied electrified solid-liquid interface. Otani and colleagues introduced an effective screening medium (ESM) method to explore half-reactions at electrode–electrolyte interfaces [15]. Their findings indicate that the applied bias potential can significantly affect the properties of water layers near the surface. The water hydrogen-bond network is found to be incredibly disruptive at negative potentials. These potential changes force the water molecules to reorientate to accommodate the new electrostatic forces. Cheng et al. developed a computational standard hydrogen electrode (cSHE) method to understand the mechanisms of interface water molecules used to control the potential of zero charge (PZC) [16, 17]. The PZC is the potential at which the electrode surface is electrically neutral, and it provides valuable information about the electrochemical properties of the system. Their study revealed that the double layer capacitance shape is a result of the interaction between the water and the surface. The orientation of the adsorbed water molecules can greatly influence the interfacial properties. Bouzid and Pasquarelllo developed a constant Fermi-level AIMD method to study the Pt(111)-water interface with the consideration of variable electrode potential referenced to the SHE. They observed significant double-layer reorganization dependent on the applied bias potential and accurately estimated the double-layer capacitance as well as the PZC of the metal electrode [18, 19]. Furthermore, Groß et al. employed AIMD with the RPBE-D3 functional to investigate the Pt(111)/water interface [20, 21, 22]. Their study demonstrated the importance of accounting for van der Waals corrections since water molecules exhibited different symmetry and density with and without van der Waals corrections. While AIMD-based simulations have been successful in studying the Pt(111)/water interface, the computational expense of AIMD limits its applicability for large-scale investigations of different solid-liquid interfaces, highlighting the need for the development of computationally efficient yet accurate methods.

To this end, Taylor and colleagues developed a static density functional theory (DFT)-based technique that enables the calculation and the manipulation of the electrostatic potential in a two-phase system, providing a cost-effective approach for studying electrified solid interfaces [23]. To calculate the applied electrode potential relative to the Fermi level ($E_{Fermi}$), a thin vacuum layer is incorporated within the solution region as an internal reference potential, allowing for comparison to the standard hydrogen electrode (SHE) with an absolute reduction potential of 4.44 V at 298.15 K [24]. However, this artificial inclusion of a vacuum layer does not account for the nature of the vacuum/solution interface. To address the limitations of including an artificial vacuum layer in the simulation, a model that does not require the presence of a vacuum layer is sought, allowing for a more accurate representation of the solid-liquid interface.

A recent study by Haruyama et al. proposed an implicit solvation method to determine the electrode potential in a system comprising an electrode and an electrolyte solution [14]. In their model, the solvent is treated as a continuous medium, taking into account its collective effects on the electrode-electrolyte interface [25, 26]. By immersing the solute in a solvent "bath" and incorporating the averaging over the solvent's degrees of freedom, the simulation approach enables the study of the electrochemical double layer and allows for the manipulation of the system's potential [27-30]. However, the results obtained from *ab-initio* molecular dynamics (AIMD) simulations highlight the significance of including explicit water layers to accurately describe the interfacial properties of the system [15-17]. Therefore, the simulation using an implicit solvation model alone may not be sufficient, as evidenced by the calculated potential of zero charge (PZC) of Pt(111) being significantly higher than the experimental value [21].

Recently, Xu et al. developed an affordable grand-canonical density functional theory (GC-DFT) approach with acceptable accuracy, which can provide an approximation of reality of the widely studied Pt(111)-water interface [31, 32]. To accurately capture the solvent environment, they employed a hybrid approach that includes both explicit and implicit solvation. This hybrid approach allowed us to capture the essential features of the electrified interface while maintaining computational efficiency. Despite its simplicity, the proposed model serves as a valuable starting point for understanding solid-liquid interfaces. This method has been successfully validated by comparing the calculated PZC and capacitance with experimental data [31, 32], demonstrating its accuracy and reliability in capturing the electrochemical properties of the electrified interface. However, only small potential window was considered in this study. As a result, the linear relationship between the applied potential and the surface charge was proposed, which was different from the previous AIMD results with a larger potential window [33]. Moreover, the adsorption of hydrogen atoms on the electrified Pt(111) surface is known as the Volmer step for HER, which largely determines the energy conversion efficiency of this important electrocatalytic reaction for hydrogen production [34, 35]. Understanding the changes in electrochemical properties and the effects of H adsorption on the Pt(111) surface is valuable for elucidating the mechanisms of electrochemical reactions and developing improved catalysts for hydrogen-related applications. However, the impact of adsorption hydrogen atom on the interfacial properties are unclear.

In this study, we used the GC-DFT method with the hybrid solvation model to reinvestigate the electrified Pt(111)/water interface. Our results reproduced the relationship between the applied bias potential and the surface charge found in the AIMD simulations. The subtle impact of the adsorbed hydrogen atom on the properties of the electrified interface was also revealed.

## 2. Computational Details

In this study, the Vienna Ab-initio Simulation Package (VASP) was utilized for all computations, employing the projected augmented wave (PAW) method [36 37]. The electron-ion interaction was described using the PAW pseudopotentials, where the valence electronic configurations of Pt, H and O are $6s^1 5d^9$, $1s^1$, and $2s^2 2p^4$, respectively. The kinetic energy cut-off value was set to 600 eV to expand the smooth part of the wave function, as suggested in previous studies [26]. During structural optimizations, all atoms relax until the Hellmann–Feynman forces were smaller than $5 \times 10^{-2}$ eV/Å. The convergence criterion for the electronic self-consistent loop was set to $1 \times 10^{-5}$ eV. The gamma-centered k-point meshes were generated with a reciprocal space resolution of $2\pi \times 0.04$ Å$^{-1}$. The optPBE exchange-correlation (XC) functional was chosen for this study as it showed success at predicting solids and water properties [38, 39]. The optPBE functional incorporates nonlocal correlation contributions into the XC energy, allowing for the consideration of less spherical densities and providing a better approximation of weak interactions. [38, 40, 41].

The Pt(111) surface was presented by a four-layer $(2 \times \sqrt{3})$ Pt(111) slab. Based on previous studies, it has been shown that a single explicit water layer is sufficient to capture the characteristics of the inner Helmholtz layer [42]. Therefore, one layer of explicit water bilayer was placed on both sides of the Pt slab here, while an implicit water layer with a thickness of 20 Å was also included. Each explicit water bilayer was composed of eight water molecules arranged in a honeycomb configuration. The implicit continuum solvation model was implemented using the VASPsol package, with a permittivity value of 78.40 and a Debye length of 3.040 Å [25, 26].

To simulate an applied bias potential (U), the double layer charge of the system was varied [43]. The structures were then further fully optimized. The applied bias potential was derived from the Fermi energy of the interface systems ($E_{Fermi}$) with respect to the SHE potential using a double-reference method, as illustrated in **Fig. 1**. In this approach, the Fermi level is first referenced to the electrostatic potential in the center of the electrolyte, which is systematically shifted to zero ($E_{shift}$), as required from the Poisson-Boltzmann equation. Then, the electrostatic potential of the implicit solvent is further referenced to the SHE potential using an additional empirical alignment term of 4.6 V, which was obtained as the difference between the experimental and computed PZC values using the implicit solvation method [26]. Hence, the applied bias potential in terms of the SHE (*U(V) vs. SHE*) can be calculated as below:

$$U(V)\ vs.SHE\ =\ -(E_{Fermi}+E_{shift})/|e|-4.6$$

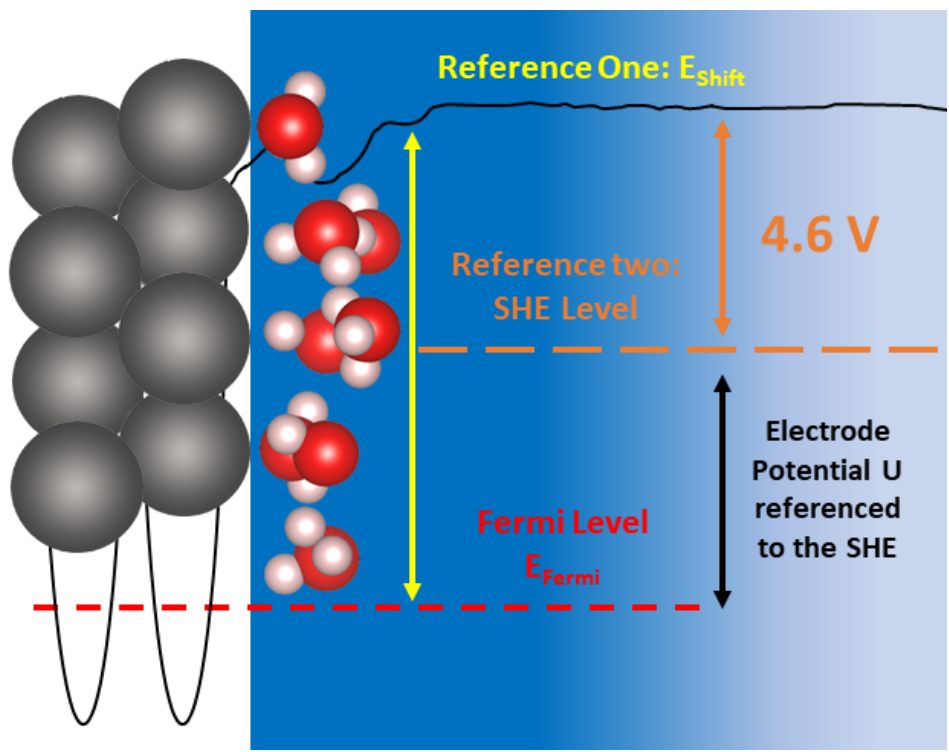

**Figure 1.** Illustration of the double reference method for calculating the applied bias potential in terms of the SHE potential from the Fermi level of the system (black: Pt, red: O, and pink: H). The average electrostatic potential in the center of the implicit solvation model ($E_{shift}$) is used as the first reference, and the SHE potential is used as the second reference.

To better understand the interaction between water and the surface, we calculated the dipole orientation between the bisector of the hydrogen bonds of water and the surface normal [32]. Therefore, the average distance of the water layer to the Pt surface as well as the bond angle between OH bond and the normal of the surface were calculated. The average distance between the Pt surface and the water layer was determined by calculating the average distance between the z positions of the oxygen atoms in the water molecules and the outmost Pt layer. The bond angle was calculated as follows,

$$\theta = arccosine\left(\frac{Z(H) - Z(O)}{d_{O-H}}\right)$$

where $Z(H) - Z(O)$ is the difference between z coordinates of H and O atoms with the vector from O towards H. And $d_{O-H}$ is the corresponding O-H bond length in water. This value can then be used to calculate the dipole orientation of the water molecule.

Three sites were considered for H adsorption (see **Fig. S1**). It was found that the atop site is the most energetically preferred one, which agree with previous conclusions [32, 44]. Consequently, only atop site will be considered here. The adsorption energy of hydrogen atom ($E_{ads}$) was calculated as below:

$$E_{ads}(V) = E_{Total}(V) - E_{Clean}(V) - E_H$$

Here, $E_{Total}$ is the total energy of the interface with adsorbed H atom at a specific electrode potential, $E_{Clean}(V)$ is the total energy of the pristine interface at a specific electrode potential, and $E_H$ is the half of the energy of a H$_2$ molecule in an isolated cell with the continuum method.

The hydrogen binding free energy was also calculated to better understand the binding properties. The $\Delta G_{H*}$ was calculated as below:

$$\Delta G_{H*} = \Delta E_{ads} + \Delta ZPE - T\Delta S$$

Here, ZPE is the zero point energy [45], T is temperature according to the recent experiment and $\Delta S$ entropic contributions retrieved from the literature [45, 46].

## 3. Results and Discussion

### 3.1 Atomic interface model

From **Fig. 1**, the double-reference model was employed to compute the properties of the Pt(111)-water interface. The interface was surrounded by a dielectric continuum representing the average electrostatic attraction of the water molecules. The adequacy of this solvation model has been a subject of prolonged debate within the scientific community. Bramley et al. and Akinola believed that the DFT-based implicit solvent is problematic [47-49]. However, several groups also suggested that the solvation model can effectivley predict interfacial properties [14-17, 50]. **Fig. 2a** shows the calculated electrostatic potential of the Pt(111) in the absence of an explicit water layer but in the presence of the implicit solvation model. The calculated PZC value obtained is 1.09 V vs SHE, which significantly differs from the experimentally reported value of 0.4 V. [51-55]. Our calculated PZC value from the implicit model is close to a similar work by Mathew et al. [21], which is 0.85 V. The small difference may be ascribed to the use of different XC functionals. Our findings support the conclusion made by Mathew et al. that the implicit solvation model alone is inadequate for accurately reproducing experimental observations [21].

Subsequently, an explicit water layer was added to the top side of the Pt(111) surface, resulting in an asymmetric system. However, the electrostatic potential of the system exhibits poor convergence in the bulk solvent region, as indicated by the inset of **Fig. 2b**. To address the convergence issues, one possible approach is to use a very thick implicit solvation layer, although this would significantly increase computational cost. Another method is the dipole correction, which helps counterbalance the slow convergence with respect to the size of the supercell [56]. However, the dipole correction method is not suitable for charged systems, and currently, it can only be applied to cubic supercells in the VASP code implementation.

In order to address the asymmetry of the electrostatic potential, we employed a "sandwich" model by mirroring the system along a perpendicular plane. This symmetric configuration enables rapid convergence of the electrostatic potential in the implicit water region, shown in **Fig. 2c**. The use of dipole correction becomes not necessary in this case, allowing for the investigation of charged systems. To this end, only the sandwich model, which showed improved convergence and symmetry in the electrostatic potential, will be utilized for further investigations in this study. The combination of explicit and implicit solvation methods rectifies limitations inherent in both approaches. This not only results in a more accurate depiction of water-water interactions in proximity to the surface but also considers water orientation.

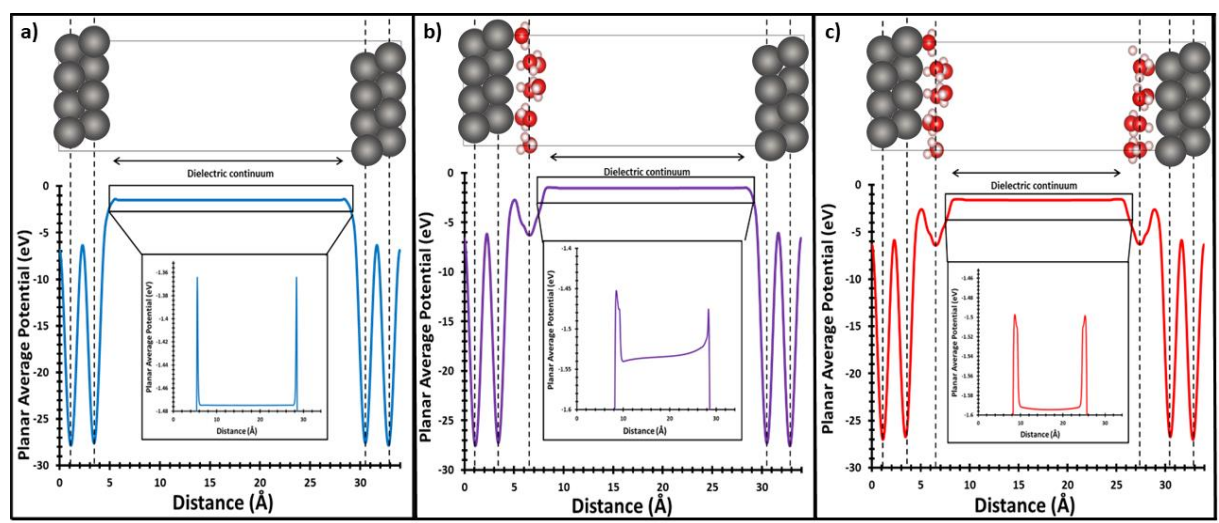

**Figure 2.** (a) The side view of the electrified Pt(111)-water interface using the implicit-only solvation model along with its planar average electrostatic potential; (b) the side view of the asymmetric electrified Pt(111)-water interface model with one explicit water layer on the top side of Pt(111) along with its planar average electrostatic potential; and (c) the side view of the symmetric electrified Pt(111)-water interface model with one explicit water layer on both the top and bottom sides of Pt(111) along with its planar average electrostatic potential. The insets amplify the changes in the electrostatic potential around the center of the implicit solution, providing a closer look at the variations in potential within that region. (black: Pt, red: O, and pink: H).

## 3.2 Water orientation

Various initial configurations were assessed, each featuring distinct water orientations. These configurations were labeled based on the orientation of OH bonds pointing away or towards the surface. These models included configurations where the hydrogen in water pointed away from the surface (H-up), towards the surface (H-down), and a mixed orientation (H-mix) configuration (see **Fig. 3**). The H-down orientation exhibited the best performance among the studied models, as it showed the closest agreement with experimental results in terms of PZC values. The experimental potential of zero charge (PZC) is around 0.4 V vs SHE [51-55]. Our PZC with the H-down configuration is 0.44 V vs SHE. Additionally, the water bilayer with the H-mixed and H-up configurations have a much lower PZC values of -0.01 and -0.50 V, respectively. Hence, even minor modifications to the explicit water layer in this interface configuration can yield substantial alterations in the overall interface properties, consistent with the findings of Xu et al. [31]. However, there are slight deviations between our calculated PZC values and the values reported by Xu et al., which were measured at 0.23 V vs. SHE [31]. This disparity can be attributed to various factors. In their research, the D3 vdW correction method and a low energy cut-off of 400 eV were employed. Additionally, their study incorporated a difference of 4.44 V between the electrostatic potential of the implicit solvent and the SHE potential.

The H-down orientation the most energetically stable one, which total energy is 0.04 and 0.10 eV lower than that of H-Mix and H-Up (see **Table. S1**). The H-down water orientation configuration was also demonstrated as the most energetically favorable and most likely to be observed at the interface by several groups from previous simulations [17, 31-33, 44]. It further supports that the H-down configuration is the dominant one in the Pt(111)-water interface.

Building upon these findings, we proceeded to conduct calculations on the H-down configuration for the remainder of the study.

The calculations demonstrate the essential role of water molecules in the model for accurately predicting the electrochemical properties of the interface. The inclusion of water molecules allows for a more realistic representation of the system, leading to improved agreement with experimental observations. Without considering the presence of water, the predictions may deviate significantly from the actual behavior of the electrified interface. Indeed, the orientation of water molecules and hydrogen bonding play a crucial role in electron transfer and intermolecular interactions at the electrified interface [57]. Implicit solvent methods, which approximate the solvent as a continuous medium, often struggle to accurately capture these short-range interactions, including hydrogen bonding [58, 59]. The explicit inclusion of water molecules in the model allows for a more detailed description of the intermolecular interactions, enabling a more accurate prediction of electron transfer processes and other properties of the electrified interface. Based on our results, further investigations in this work will focus on the H-down configuration as it represents the most dominant configuration for the electrified interface.

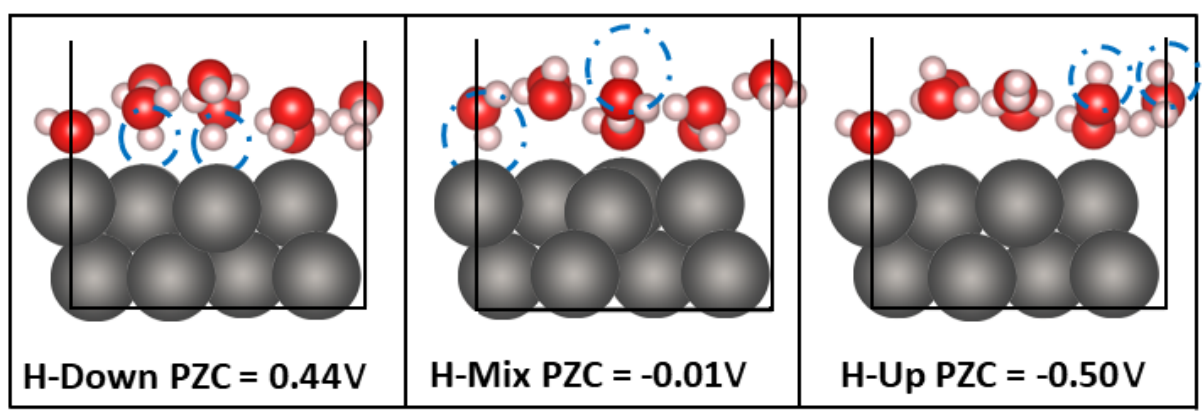

**Figure 3.** Optimized atomic structures of three Pt(111)-water interface with specific water orientations with respect to the Pt(111) surface. (black: Pt, red: O, and pink: H).

The impact of the thickness of the implicit solvation layer on the PZC values was investigated, considering implicit layers of 20, 25, and 30 Å thickness (see **Fig. S2**). The calculated PZC values were found to be nearly identical, indicating that a thickness of 20 Å for the implicit solvation layer is sufficient to obtain converged properties of the electrified Pt(111)-water interface.

### 3.3 Double Layer Capacitance

The computation of the double layer capacitance based on our models revealed two linear trends in the evolution of double layer charge vs. electrode potential, as shown in **Fig. 4a**. The capacitance values are obtained by taking the derivative of the surface charge with respect to the potential. In **Fig. 4b**, a finite difference approach was used to calculate the derivative between two adjacent points, resulting in an average capacitance within that range. A small potential range from -1 to 1 V vs. SHE was considered, which is wider than that used in previous studies [31, 32, 44]. Small ranges are often used in both computational and experimental studies to prevent unwanted electrochemical reactions such as electrolysis [5]. Additionally, one water layer would likely be insufficient to describe the interaction at higher potentials [5].

Our results suggest a change in the capacitance at around 0.5 V. In the lower potential region (< 0.5 V) around the PZC, the double layer capacitance was calculated to be 21.64 $\mu F/\text{cm}^2$, which agrees well with experimental measurements of about 20 $\mu F/\text{cm}^2$ [53], while higher potential range (> 0.5 V) presented a different capacitance of 46.59 $\mu F/\text{cm}^2$. **Fig. 4b** illustrates the relationship between the applied potential and the capacitance, and it demonstrates a close reproduction to the experimental measurements [33, 60]. This indicates that our computational models effectively capture the behavior of the capacitance as a function of the applied potential. It is worth noting that the capacitance change shown in **Fig. 4b** is noisy, which was expected as this approach is oversimplified and neglecting the dynamic feature of the water.

Additionally, electrolytes are not explicitly described here. And electrolyte adsorption and the formation of the outer Helmholtz plane are essential components of the electrified solid liquid interface at these potentials [5, 61], which may lead to the theory-experiment discrepancy observed in **Fig. 4b**.

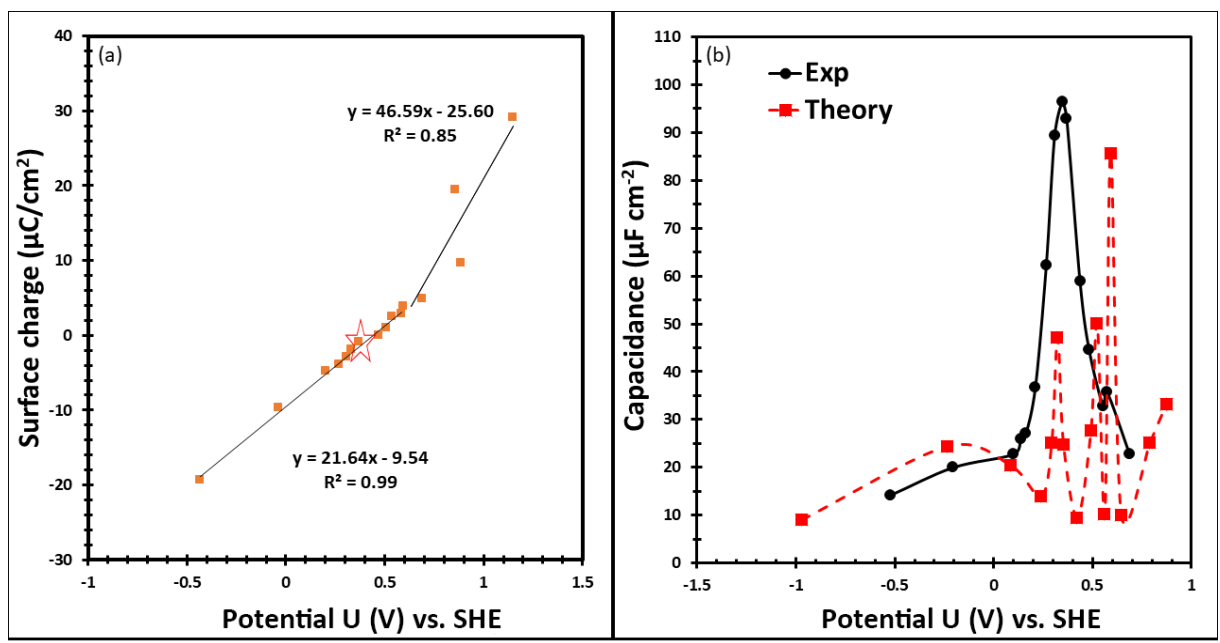

**Figure 4.** (a) The double layer charge as a function of the calculated applied bias potential. The star indicates the PZC. (b) the calculated double layer capacitance as a function of the applied bias potential along with the experimental data, which are retrieved from [53, 62].

### 3.4 Structural properties

The structural properties of the interface under an applied bias potential play a crucial role in the initial steps of electrochemical processes, including oxygen evolution reaction (OER), hydrogen evolution reaction (HER), and carbon dioxide reduction reaction (CO2RR). In this study, the water position to the surface provides information about the proximity of the water molecules to the Pt(111) surface. It shows how the distance between the water layer and the surface changes with the applied potential, reflecting the interaction between the water molecules and the electrified surface. The average dipole orientation of the water molecules with respect to the surface normal indicates the orientation and arrangement of the water molecules at the interface. It provides insights into the hydrogen bonding network and the structural properties of the water layer in the presence of the electrified Pt(111) surface.

**Fig. 5** shows the change of the water height and dipole orientation as a function of the applied bias potential. Our findings indicate that the double layer charge of the interface exhibits two distinct linear domains, which are influenced by the distance between the metal and the water layer as well as the orientation of the water molecules. At positive potentials, the distance between the water and metal interface decreases with increasing potential, while it increases at negative potentials. Furthermore, the analysis of the dipole orientation indicates that water molecules point away from the interface under positive potentials. This behavior can be attributed to the electrostatic interaction between the positively charged surface and the negatively charged oxygen atoms, causing the hydrogens to be pushed aside. These observations are aligned with findings from other studies [17, 32]. Under positive potentials, an interesting phenomenon is observed where water molecules approaching the surface can experience collisions with each other. At a potential of 0.5 V, a majority of the water molecules are adsorbed on the surface in an H-up configuration. This leads to a disruption and decreased organization of the water layer. In contrast, at lower potentials, a more rigid ice-like layer forms as water molecules move away from the surface. The height variance of all water molecules was calculated to assess the degree of spatial organization (see **Fig. S3**). Variance is a statistical measure that quantifies the dispersion of a data set around the average. Using this variable, the organization of the water molecules can be quantified. At a potential of 0.2 V, the height variance of water was found to be 0.37, indicating a relatively lower level of spatial organization. However, at 0.7 V, the variance increased to 0.42, suggesting a higher degree of disorder and less spatial organization within the water layer. This change in variance further supports the observation that the water layer becomes disrupted and less organized at higher potentials. These observations highlight the dynamic behavior of the water layer and its dependence on the applied potential, resulting in changes in the overall structure and organization of the interface.

The orientation of water molecules at the electrified interface indeed plays a crucial role in determining the electrochemical properties. Specifically, at lower potentials $H_2O$ contribute to the formation of hydrogen bonding networks that enhance the stability and organization of the water layer, which leads to a denser and more organized water layer [63]. The enhanced stability and organization of the water layer in these orientations can affect various electrochemical processes. Xu, et al. employed a comparable methodology and yielded outcomes consistent with our findings [31]. Their observations also highlighted the pronounced influence of water structures on interface behavior. This congruence underscores the reproducibility of our approach and its potential applicability in other research areas.

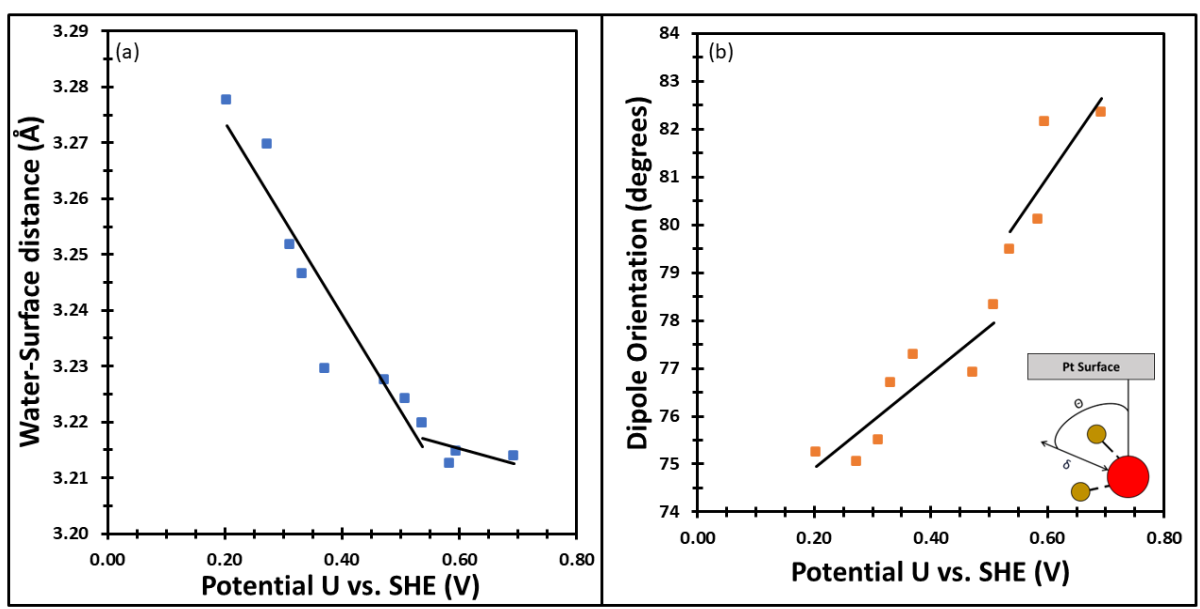

**Figure 5.** (a) Water-surface distance and (b) Average dipole orientation of the water molecules respect to the surface normal as a function of applied bias potential.

**3.5 H adsorption**

The influence of applied bias potentials on the adsorption properties of H atoms on the electrified Pt(111)-water interface were also investigated here. The atop site was the energetically preferred with the lowest energies (see **Table. S2**). The PZC of the system with the adsorbed H atom at the atop site is 0.38 V, which is lower than that without adsorbed H. The lower PZC value suggests that the presence of H alters the double layer charge distribution and influences the electrochemical properties of the interface. This observation is consistent with the fact that H adsorption on Pt surfaces can affect the surface chemistry and catalytic activity, particularly in the context of the HER. The aqueous hydrogen binding free energy change ($\Delta G_{H*}$) was also computed at a calculated zero-point energy change (ΔZPE) of 0.24 eV, energy change (ΔE) as -0.26 eV, and a entropy change (ΔS) at 343K obtained from Yang et al. [46]. The resulting $\Delta G_{H*}$ value was determined to be 0.18 eV, which exhibits an exceptionally close correspondence to the experimental value of 0.10 eV [46]. This convergence between the calculated and experimental values underscores the reliability of the method and its capability to predict the properties of the solid-liquid interface.

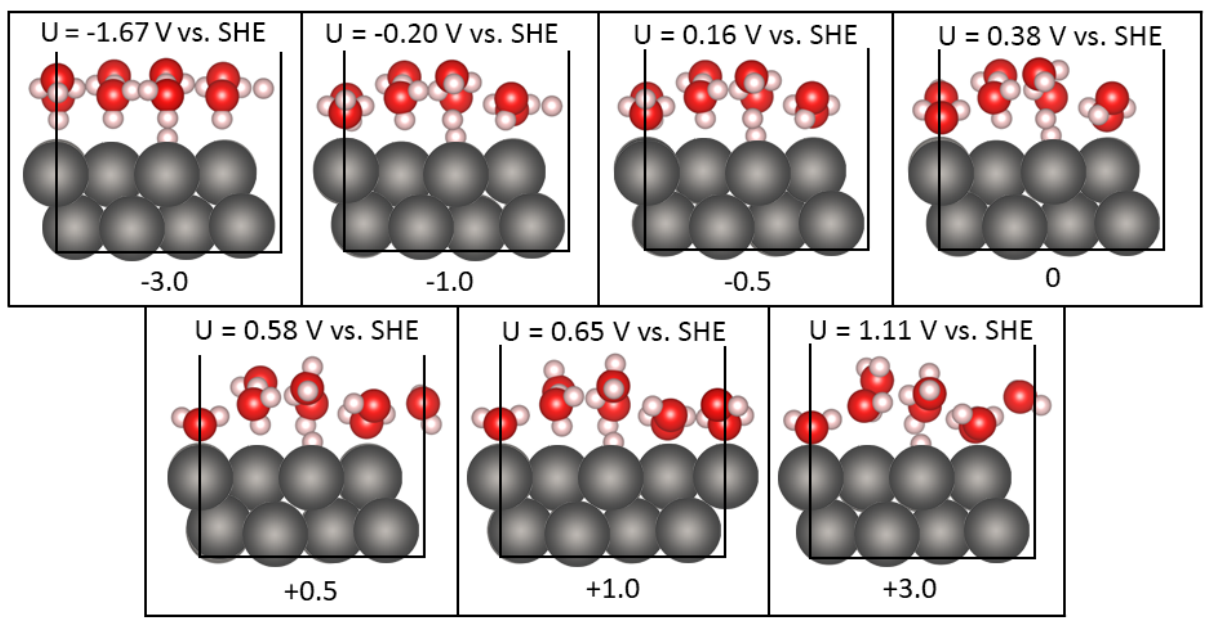

**Figure 6.** H adsorption on the electrified Pt solid-liquid interface. The number underneath the structure refers to the total charge of the system. The U (SHE) above the structure is the potential at that charge. Electrons were added or removed to give the system a charge (black: Pt, red: O, and pink: H).

Similar to the Pt(111)-water interface without adsorbed H, **Fig. 7a** shows that the double layer charge for the Pt(111) surface with adsorbed H exhibits two linear trends as the applied potentials increase. This suggests that there are two distinct regimes in the behavior of the double layer charge as a function of the applied potential. There are several distinctions caused by the hydrogen adsorption. The presence of adsorbed hydrogen on the Pt(111) surface leads to a reduction in capacitances at the low (< 0.5 V vs. SHE) and high applied bias potentials (> 0.5 V vs. SHE). The reduction in capacitance can be attributed to several factors. Firstly, the adsorption of hydrogen atoms on the surface may alter the electronic structure and charge distribution, leading to changes in the surface capacitance. Additionally, the presence of adsorbed hydrogen can influence the surface chemistry and the interaction of the electrode with the water, thereby affecting the charge storage capacity.

The adsorption energies of hydrogen on the Pt(111) surface are influenced by the applied potential. Similar to the double layer charge, the adsorption energy can be described by two linear relationships as the potential is varied (see **Fig. 7b**). At high applied potentials, the adsorption of hydrogen on the Pt(111) surface becomes significantly stronger. This can be attributed to the increase in double layer charge resulting from the higher negative potential. When the potential is raised, the surface carries more positive charge (See **Fig. 7a**). The increased double layer charge enhances the electrostatic attraction between the negatively charged O atoms in water and the positively charged surface, resulting in a more stable adsorption configuration. [32].

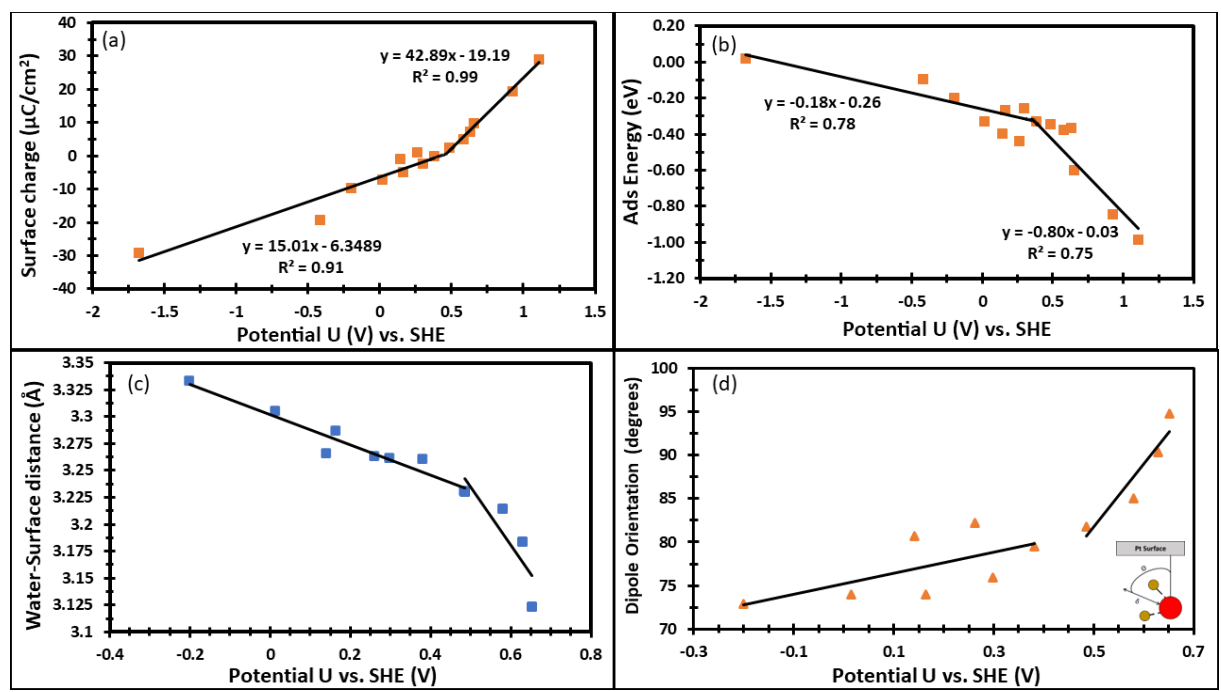

**Figure 7.** The dependence of (a) double layer charge, (b) H adsorption energy, (c) height of water from surface, and (d) average dipole orientation of the water molecules with respect to the surface normal on the applied potential when hydrogen adsorption on the electrified Pt(111) surface.

The height of water and its dipole orientation at the Pt(111)-water interface with the adsorbed H atoms closely resembled those observed on the interface without adsorbed H atoms (see **Figs. 7c** and **d**). The water height and dipole orientation of the water molecules varied depending on the applied potential with two distinct linear domains. At the point of zero charge, the presence of the protruding H atom at the Pt(111)-water interface caused a bending effect on the water layer. This occurred because neighboring water molecules repelled each other through hydrogen bonding interactions. As the applied potential increased, this effect became stronger, leading to a maximum height variance in the water position of 0.37 at 0.63 V. Conversely, as the potential decreased, the water molecules exhibited a tendency to move away from the surface, resulting in the formation of a flatter water layer. The variance in the distance between the water molecules and the surface decreased with decreasing potential, indicating that the water molecules were approximately located at the same height. For example, at -1.67 V, the height variance was measured to be 0.19. The dipole orientation results are also close to previous observations without adsorbed H regarding the orientation of water molecules at the Pt(111)-water interface under different potentials (see **Fig. 7d**). Under positive potentials, the orientation of the water molecule's hydrogen atom is directed away from the surface, whereas under negative potentials, it points towards the surface. A distinct disparity emerged following hydrogen adsorption. The model with adsorbed hydrogen exhibited a broader dipole range compared to the clean surface, and the distance between water and the surface was greater. The presence of adsorbed hydrogen amplified the trends observed on the clean surface (see **Fig. 4-7**). Therefore, the presence of the adsorbed hydrogen species influences the behavior of the water molecules. The orientation analysis reveals that the water molecules adjacent to the adsorbed H atoms adjust their positions to accommodate the adsorbed hydrogen into the hydrogen bonding network. The adsorbed H atom prevents the neighboring water molecules from approaching too close to the surface.

Interestingly, despite the observed changes in adsorption strength with applied potential, the bond length between the adsorbed H atom and surface Pt atom remained relatively constant at approximately 1.5 Å. Given that the trends in adsorption energy and water height show a striking similarity, it indicates that the adsorption energy is influenced not only by the Pt-H bond but also by the interactions between water and the adsorbed hydrogen. The strong adsorption energy at positive potentials can be attributed to the integration and support of the adsorbed hydrogen species within the water network. In contrast, at low potentials, the water molecules move away from the surface with H atoms in water pointing downwards, leaving no available positions for the adsorbate to interact with the water network. As a result, the adsorption energies change more slowly along the applied potential. These findings highlight the intricate interplay between the adsorbed hydrogen, water molecules, and the electrochemical environment, which affects the adsorption behavior and overall properties of the Pt(111)-water interface.

Nevertheless, it is essential for future investigations to expand the scope of this model by integrating other solvent parameters like electrolytes and pH. Furthermore, while our current study has primarily examined the interface under low potentials to avoid undesired chemical reactions, it's crucial to acknowledge that many heterogeneous electrocatalytic reactions necessitate higher voltages. In such instances, a single water layer might not suffice for accurately modeling these electrified interfaces. The consideration of water coverage and the establishment of inner and outer Helmholtz planes are likely imperative.

## 4. Conclusion

In conclusion, we used an affordable approach based on the GC-DFT method to reinvestigate the electrified Pt(111)-water model system. We used a hybrid approach with a "sandwich" design to combine explicit and implicit solvation methods, which can overcome convergence issues and provide more reliable results. The relationship between electrochemical properties including double layer charge and capacitance and the applied bias potential derived from our GC-DFT results shows correlation with two distinct linear domains, in good agreement with experimental measurements. The great match highlights the importance of considering explicit water molecules and their interactions in modeling electrified solid-liquid interfaces accurately. Further analysis on the structural properties of the interface demonstrates that the orientation and position of water molecules play a significant role in determining the PZC and double layer capacitance of the system. At high potentials, the water layer becomes disordered and moves away from the surface, while at low potentials, the water molecules point towards the surface, stabilizing the structure. The adsorption of hydrogen on Pt also affects the system's performance, with different adsorption energies observed at different potentials.

This affordable GC-DFT approach provides a valuable and efficient means to enhance our understanding of electrified solid-liquid interfaces. By combining the advantages of implicit and explicit solvation methods, this approach offers an accurate description of the interactions and properties of these interfaces while minimizing computational costs. The agreement between the findings of this approach and those reported in AIMD simulations further validates its effectiveness. Overall, the affordable GC-DFT approach enables researchers to study electrified solid-liquid interfaces in a cost-effective manner without compromising accuracy, contributing to our understanding of these important systems.

**Acknowledgments:** This research was conducted on the supercomputers in National Computational Infrastructure (NCI) in Canberra, Australia, which is supported by the Australian Commonwealth Government, and Pawsey Supercomputing Centre in Perth with the funding from the Australian government and the Government of Western Australia. Calculations were performed by using resources from Grand Equipement National de Calcul Intensif (GENCI, grant no. A0120913426). Computational resources provided by the computing facilities Mésocentre de Calcul Intensif Aquitain (MCIA) of the Université de Bordeaux and of the Université de Pau et des Pays de l'Adour. We acknowledge funding from CNRS Institute of Chemistry through the "International Emerging Actions 2022" mobility grant (2DH2 project) and from Agence national de recherche under the RECIFE ANR-DFG project (Grant Number ANR-21-CE08-0036-01).

**Supplementary Information**

Additional supporting information can be found attached to accommodate the manuscript.